\documentclass[10pt, conference]{IEEEtran}
\usepackage[ruled,vlined]{algorithm2e}
\usepackage{cite,url}
\usepackage{amsmath,amssymb,amsfonts}
\usepackage[]{algorithm2e}
\usepackage{graphicx}
\usepackage{textcomp}
\usepackage{xcolor}
\usepackage{balance}
\usepackage{makecell}
\usepackage{ctable}
\usepackage{tabularx,booktabs}
\def\BibTeX{{\rm B\kern-.05em{\sc i\kern-.025em b}\kern-.08em
    T\kern-.1667em\lower.7ex\hbox{E}\kern-.125emX}}
\usepackage{multirow}
\usepackage{tabularx}

\makeatletter
\def\thickhline{%
  \noalign{\ifnum0=`}\fi\hrule \@height \thickarrayrulewidth \futurelet
   \reserved@a\@xthickhline}
\def\@xthickhline{\ifx\reserved@a\thickhline
               \vskip\doublerulesep
               \vskip-\thickarrayrulewidth
             \fi
      \ifnum0=`{\fi}}
\makeatother

\newlength{\thickarrayrulewidth}
\newcommand{\reff}[1]{Fig.~\ref{#1}}
\newcommand{\reft}[1]{Table \ref{#1}}

\begin{document} 
\title{Gateway Station Geographical Planning for Emerging Non-Geostationary Satellites Constellations}
\author{\IEEEauthorblockN{Victor Monzon Baeza, Flor Ortiz, Eva Lagunas, Tedros Salih Abdu, and Symeon Chatzinotas}
\IEEEauthorblockA{\textit{Interdisciplinary Centre for Security Reliability and Trust (SnT), University of Luxembourg, Luxembourg}\\
\textit{Email: \{victor.monzon, flor.ortiz, eva.lagunas, tedros-salih.abdu, symeon.chatzinotas\}@uni.lu}} }
  
\date{}
\maketitle

\begin{abstract} 
Among the recent advances and innovations in satellite communications, Non-Geostationary Orbit (NGSO) satellite constellations are gaining popularity as a viable option for providing widespread broadband internet access and backhauling services.  However, a more complex ground segment with multiple ground stations is necessary due to these satellites' high speeds and low altitudes. The complete dimensioning of the ground segment, including  gateway optimal placement and the number of ground access points, remains a relevant open challenge. In this article, we provide an overview of the key factors that shall be considered for NGSO gateway station geographical planning. Subsequently, we propose a ground segment dimensioning approach that combines several criteria, such as rain attenuation, elevation angle, visibility, geographical constraints, and user traffic demands. The operational concept is first discussed, followed by a methodology that combines all these constraints into a single map-grid to select the best position for each gateway. Furthermore, a case study is presented, which demonstrates the performance of the proposed methodology, for one example constellation. Finally, we highlight relevant open challenges and key research directions in this area. 
\end{abstract}


\section{Introduction}
\label{intro} 
 
Satellite communications (SatComs) have regained attention due to recent advancements in technology and private investments, such as the Non-Geostationary Orbit (NGSO) satellite mega-constellations created for broadband communication services \cite{Hayder_Survey}. A new wave of lower-orbit SatCom systems is in the making, such as the IRIS2 constellation, embracing the benefits of lower radiation exposure and reduced latency to provide Internet access to under-served regions, which require a global, scalable, flexible, and resilient solution \cite{RIoT}. Achieving worldwide internet coverage using low-altitude and fast-moving satellites presents certain technical obstacles and challenges on the NGSO SatComs system's ground segment. 

Several ground stations need to be distributed throughout the Earth's surface to guarantee global connectivity with the components of the mega-constellations. These stations, called Gateways (GW), connect the satellite to the ground via feeder links. The GW, in the absense of Inter-Satellite Links (ISL), must always maintain visibility with some element of the constellation to offer the service continuously, which is challenging in the case of NGSO since the satellites are moving, displaying a dynamic situation. Therefore, geographical planning of the GW elements is quite challenging for NGSO. A conservative approach in which more GWs than necessary are distributed geographically can lead to oversizing, resulting in high operator costs and generally becoming economically impractical. This makes it impossible for small operators to enter the market. To reduce the necessary number of GW, ISLs are currently being introduced, which allows a communication link between each satellite and the neighboring one, thus reducing the need for all satellites to be visible by (at least) one GW. However, this is a hot-line of research even at an early stage \cite{RIoT}. Weather conditions are another limiting factor when it comes to locating a GW since feeder links usually operate on high spectral bands, which are very sensitive to weather impairments like rain fading. 

Other factors influence channel modeling compared to its geostationary (GEO) counterpart, such as the Doppler effect, which would complicate the control and visibility of Low Earth Orbit (LEO) satellites due to movement. Several channel models are presented in \cite{Victor_channel} that reflect the variability and lack of consensus in a channel model that allows us to design a unique or standard ground segment.

A preliminary analysis of ground segment requirements to face the new space age is presented in \cite{PreliminaryGS,Optimal_Strategies_GW} for the South American region. However, the GW locations are assumed to be given in \cite{PreliminaryGS}. Considering the available works in the literature, we found the need to identify the needs, gaps, and issues in dimensioning the ground segment of emerging NGSO systems. We start this paper by providing an overview of the key factors that shall be considered for NGSO gateway station positioning. Next, we discuss a novel methodology to combine different criteria such as rain attenuation, elevation angle, visibility, geographical constraints, and user traffic demands. The operational concept can be adapted to any input NGSO constellation and criteria. Finally, we present an overview of future research lines and open technical challenges.
 

\begin{table*}[]
\caption{Ground Segment Characteristics and Challenges}
\begin{tabular}{|l|l|}
\hline
\begin{tabular}[c]{@{}l@{}}Ground Segment \\ Characteristics and Challenges\end{tabular}  & Description                                                                                                                              \\ \hline
\begin{tabular}[c]{@{}l@{}}High-Speed Mobility \\ of LEO satellites\end{tabular}          & \begin{tabular}[c]{@{}l@{}}Opting for low-latency communications involves dealing with increased satellite mobility. As the satellite constellation \\moves closer to Earth, it appears to move at higher speeds when observed from the ground. Consequently, from the \\Ground Station's viewpoint, this necessitates the deployment of numerous GW stations strategically spread across the\\globe. The objective is to ensure that at least one visible GW is available for each satellite and to manage seamless\\ handovers over time.\end{tabular}                                                                                                                                                                \\ \hline
\begin{tabular}[c]{@{}l@{}}Feeder links operational \\ frequency\end{tabular}             & \begin{tabular}[c]{@{}l@{}}Feeder links interconnect GW with satellites. In the beginning, these were supposed to operate in  Ka-band. However, \\the increasing capacity demand needs have forced the move of feeder links to higher bands (i.e. Q/V bands) to have \\access to a wider bandwidth.  One may think that a further advantage of Q/V-band is that the number of GW stations \\can be decreased because more bandwidth is available per ground station.\end{tabular}                                                    \\ \hline
\begin{tabular}[c]{@{}l@{}}Weather Impairments / \\ Link availability\end{tabular}        & \begin{tabular}[c]{@{}l@{}}Moving the feeder links to higher bands solves the bandwidth limitation, as discussed.  However, heavy fading caused \\ by rain attenuation in Q/V band necessitates using ground station diversity techniques to ensure the required availability \\ \cite{Gharanjik15}. In this sense, the more GWs, the better, as we will have diversity in the link availability and avoid waiting for the \\ satellite to pass over a specific location.\end{tabular}                                                                                                                     \\ \hline
Cost of FW Deployment                                                                     & \begin{tabular}[c]{@{}l@{}}The cost of GS scales with the number of ground stations. Building ground stations (especially for Q/V band),\\ operating, and maintaining them is expensive and requires many resources, including equipment (i.e., antennas, modems),\\ land, and dedicated highly-qualified personnel. According to industry experts \cite{PerspectivesGS}, the cost of GS can reach one-third of \\the total cost for large programs and can represent between 10-15\% of satellite operators’ OPEX.\end{tabular}                                                                        \\ \hline
\begin{tabular}[c]{@{}l@{}}Proximity to secure \\ terrestrial infrastructure\end{tabular} & \begin{tabular}[c]{@{}l@{}}Another aspect to be considered when designing the GS for NGSO systems is the proximity of GW to terrestrial\\ point-of-presence and to points of interest such as cloud-based services. In this sense, avoiding long and expensive\\ transport networks between remote ground stations is highly desirable, especially considering the following aspect.\end{tabular}                                                                                                                                                                                          \\ \hline
\begin{tabular}[c]{@{}l@{}}Geographical and \\ Political Constraints\end{tabular}         & \begin{tabular}[c]{@{}l@{}}It is clear that NGSO constellations require a (dense) network of ground stations potentially installed in multiple countries \\and/or Earth regions. One should avoid natural-disaster-prone  countries repeatedly affected by flooding and/or earthquakes. \\Furthermore, ground station terminals constitute an important vulnerability point as a direct access point to a satellite.\\ Therefore,  geopolitical constraints, sovereignty, and potential military conflicts must be considered when selecting ground \\segment locations.\end{tabular} \\ \hline
Fast GS Reconfiguration                                                                   & \begin{tabular}[c]{@{}l@{}}GS must be designed to dynamically reconfigure in the advent of possible link failures as fast as possible to prevent\\ service outages. Preliminary works\cite{Gharanjik15} considered weather-induced outages only, the latter using machine learning\\ mechanisms for predicting the outage events. However, there is still room for improvement to consider other\\ constraints as mentioned above.\end{tabular}                                                                                                                                                         \\ \hline
\end{tabular}
\label{Tab_criteria}
\end{table*}

\section{Ground Segment Design: Criteria}
\label{GSD}

One of the fundamental challenges of the NGSO-ground segment (GS) with global coverage is the number and availability of GW locations. \reft{Tab_criteria} provides a list of aspects that make the GS design challenging. 

The need for wider bandwidths and the increasing demand for capacity are forcing the feeder links to move to higher bands, i.e., Q/V-W bands. Even feeder links in optical bands that are at low technology readiness levels. As discussed in \reft{Tab_criteria}, the Q/V band GS design is not a trivial task. New architectures for the ground segment are required beyond the proposals for band Ka in \cite{arquitectura} to support higher bandwidth and capacity. Instead, these frequencies are impaired by higher atmospheric attenuation, such as rain, which in turn causes outages in the services.
For this reason, choosing the correct position of the GW is vital to avoid areas with high rain precipitation, among other mitigating factors for the feeder link power. To overcome service interruptions, the primary strategy used so far is site diversity, consisting of redundant GW with backup stations while switching the service in the event of an interruption. An overview of the site diversity concept and different strategies are described in \cite{Review_Site_Diversity} in two representative climatic groups: a temperate region and in a tropical climatic area. To decide the position of a GW with diversity using the rain attenuation criterion, rain prediction methods such as the one proposed in \cite{Rain_predition} have been considered.

The inconvenience of site diversity strategies would increase the development cost of the NGSO ground segment, which, as mentioned, is aggravated for current and upcoming mega-constellations, regardless of the criteria used to create the redundancy GW network (rain, traffic, access, delay). A study shown in \cite{Cost} examines the optimal selection of GW to reduce the overall installation cost while ensuring an acceptable level of outage probability based on the assumption that weather conditions at each site are independent. The interest is to reduce the number of GW that compose the GS network. The works \cite{arquitectura}, and \cite{Placement_traffic} already consider minimizing the number of GW but exclusively under a single criterion, the first for atmospheric phenomena, while the second for traffic distribution. Therefore, it is not an optimal solution to provide guarantees on the total availability of the service. Also, the character of the time-varying topology of the NGSO satellites, which determines the real-time satellite visibility, has not been considered.

On the other hand, the GW placement further affects the service coverage and access performance of the network to service demands \cite{Placement_traffic,GW_Placement_Remote}. To avoid loss of service, one can also balance the traffic between the GWs, considering a service data demand distribution. The authors in \cite{Placement_traffic} propose a GW placement method for NGSO networks that identifies the best GW locations that can balance traffic loads based on constraints such as link interference, satellite bandwidth, and the number of satellite antennas. Instead, atmospheric attenuation is not considered.

\section{Multi-Criteria Approach}
\label{system}
\begin{figure}[]%
    \centering
    \includegraphics[width=3.65in]{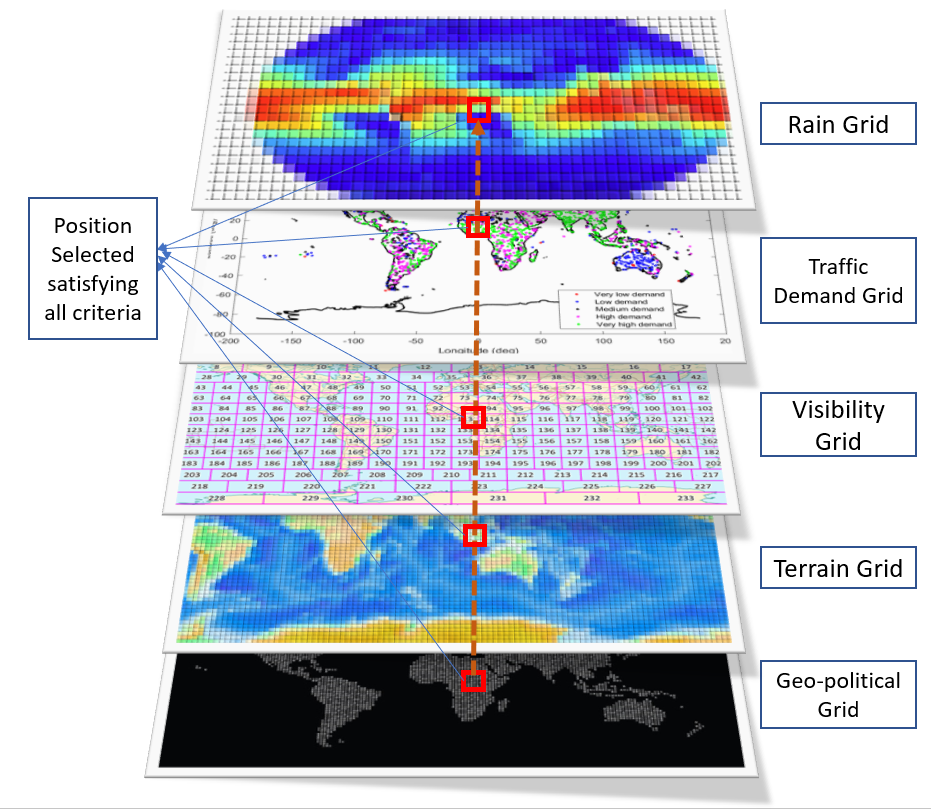} 
    \caption{Grid-model for multi-criteria approach.}
    \label{grids}
\end{figure}

This section provides a methodology to determine the best geographical positions to locate GW stations by considering multiple criteria. For this purpose, we have defined a system model based on layers, where each layer is called a \textit{grid} and represents a choice criterion, as shown in Figure \ref{grids}.

Latitude and longitude define the fundamental grid, by steps of 0.1 degrees for latitude and 0.1 degrees for longitude. The dimension of the coordinate grid determines the dimension of the rest of the grids, which has to be the same for coordinate-to-coordinate mapping. The initialization of coordinates basically depends on the internal operator restrictions in terms of regions for deployment. The selection of the step or subdivision has an important weight on the time and computation complexity required to generate the different grid levels and the overall procedure. Therefore, a balance must be found between accuracy, speed, and complexity. Each coordinate pair is a candidate position for a GW will be one that simultaneously meets the decision thresholds marked in each independent grid. The red box in \reff{grids} represents the selected coordinates that simultaneously meet all the conditions imposed to place a GW. Few of these grids have been mathematically described in our previous work \cite{EuCNC_Victor}, which did not consider the worldwide NGSO coverage area. Weather modeling is statistically represented using the ITU-R (International Telecommunication Union Radiocommunication Sector) model for quantifying rain attenuation in SatCom. Rain attenuation is calculated as a function of satellite frequency, rain rate, and geographical location. A grid with a matrix is defined to represent the weather model, where each position represents a geographical position, and the value represents the rain attenuation in dB \cite{EuCNC_Victor}.  

A population density-based traffic model estimates data demand on Satcom systems to generate traffic demand grids. Data demand is estimated using a population density-based traffic model that considers four key variables influencing data demand: the throughput per user, the population density, the penetration rate\footnote{Refers to the proportion of the population using SatComs services and is usually measured in users per inhabitant.}, and the concurrency rate\footnote{Refers to the proportion of users simultaneously using SatComs services.}. The product of these four variables gives the throughput density per square kilometer, representing the total amount of data being transmitted or received in a given area. 

Concerning visibility, two sub-grids are calculated: one for the average number of satellites that can be seen from the GW location and one for the visibility over time of a satellite from the GW location. This means that the NGSO constellation is predefined and offered as input.

The model also includes a grid to indicate the altitude concerning sea level and another to indicate for each coordinate whether or not it is allowed to place a GW for geo-political reasons.

The advantage of this multi-criteria approach is that we can include new grids according to new conditions required by each operator independently for each design. This customisation also includes the grid combination, e.g. the base grid of coordinates can be combined to present it as a triplet in which the third parameter is the station's height.  

\begin{figure*}[]%
   \centering
   \includegraphics[width=6in]{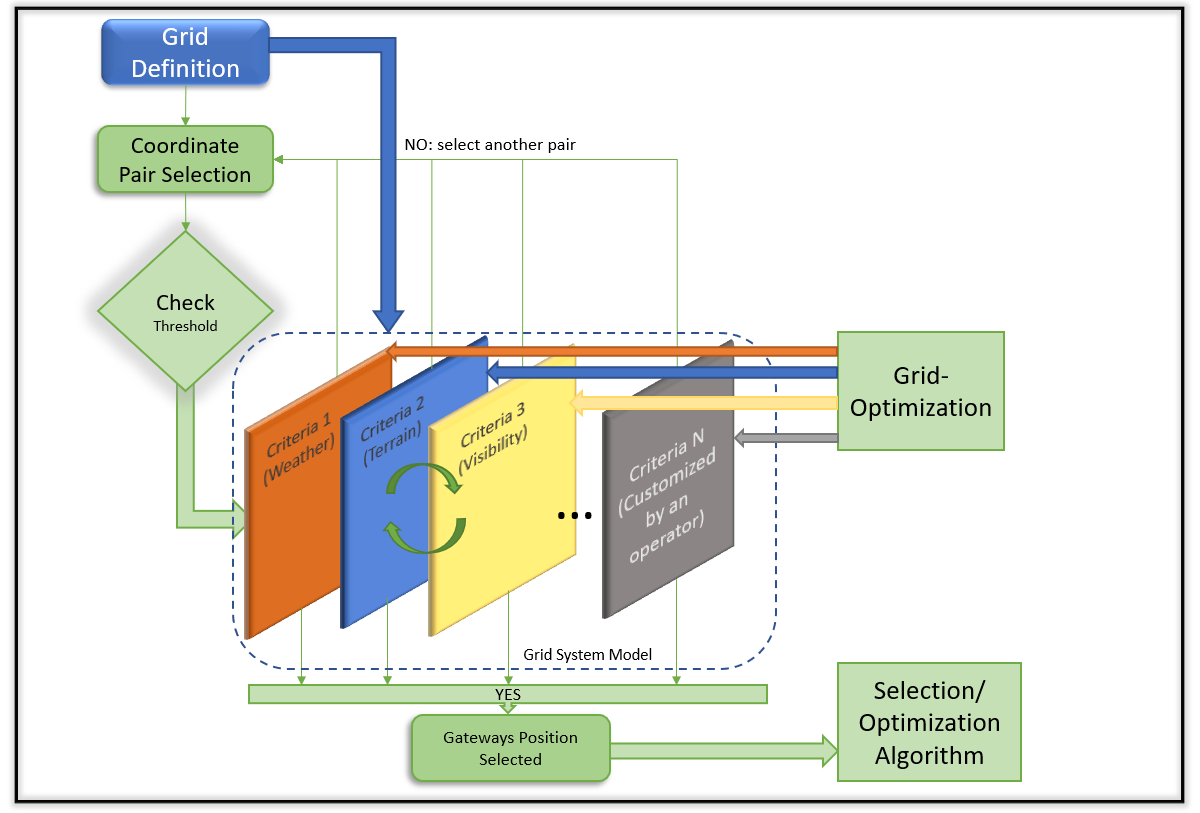} 
   \caption{Methodology and Procedure Flow.}
   \label{diagram}
\end{figure*}

To carry out the design and decide where to place the GW stations, we follow the procedure shown in \reff{diagram}. First, the grids participating in the multi-criteria decision must be designed and defined. A pair of coordinates (latitude and longitude) is selected, and each grid is traversed, checking if the value for said pair meets the threshold established for the corresponding criterion.  No information is exchanged between grids. However, multiple simultaneous grids can be considered if you wish to combine criteria. This depends on the definition of the grid.  If one of the grids returns a negative response in the check, that pair is invalid and we must select another pair of coordinates. If all the conditions are met for a pair of coordinates, we will have a candidate position to place a GW. Once all coordinates have been analyzed, we will have a list of candidate positions to place the GWs. Both the list of positions obtained and the definition of the grids may be subject to optimization at a later stage.

\section{Numerical Evaluation}
\label{results}

To evaluate the proposed multi-criteria approach, we use the criteria established in \reff{grids} (rainfall attenuation, traffic demand, visibility, geo-political constraints) for an NGSO constellation at an altitude of 800 km. The frequencies used are 19.7 GHz, 30 GHz, 40.5 GHz, and 47.2 GHz, representing both traditional and emerging spectral bands for the feeder link. In all cases, the minimum elevation angle is fixed at 10 degrees. The thresholds for the algorithm are selected to exemplify the methodology and process described here. The following thresholds per grid are defined:  

\begin{itemize} 
    \item Rain threshold (weather grid): all the rain attenuation values available in the grid have been analyzed, the maximum value selected, and based on this number, we establish 25\% attenuation as a threshold with respect to the maximum allowed attenuation.
    
    \item  Geo-political threshold: represents a binary threshold between a geographic position with conflict or not. To exemplify the proposed model in this work, conflicting positions have been distributed randomly on the map.
    
    \item  Visibility threshold: the threshold is established considering at least 3 visible satellites in each position. 
    
    \item  Traffic threshold: the threshold is established between 5 types of traffic densities, those with a high traffic density of 33 Mbps$/\text{km}^2$.
    
    \item  Terrain threshold: in this example, we have considered the threshold that determines that there is land, excluding all aquatic areas (seas, oceans, lakes, etc) 
\end{itemize}
    
\begin{figure*}[]%
   \centering
   \includegraphics[width=6.5in]{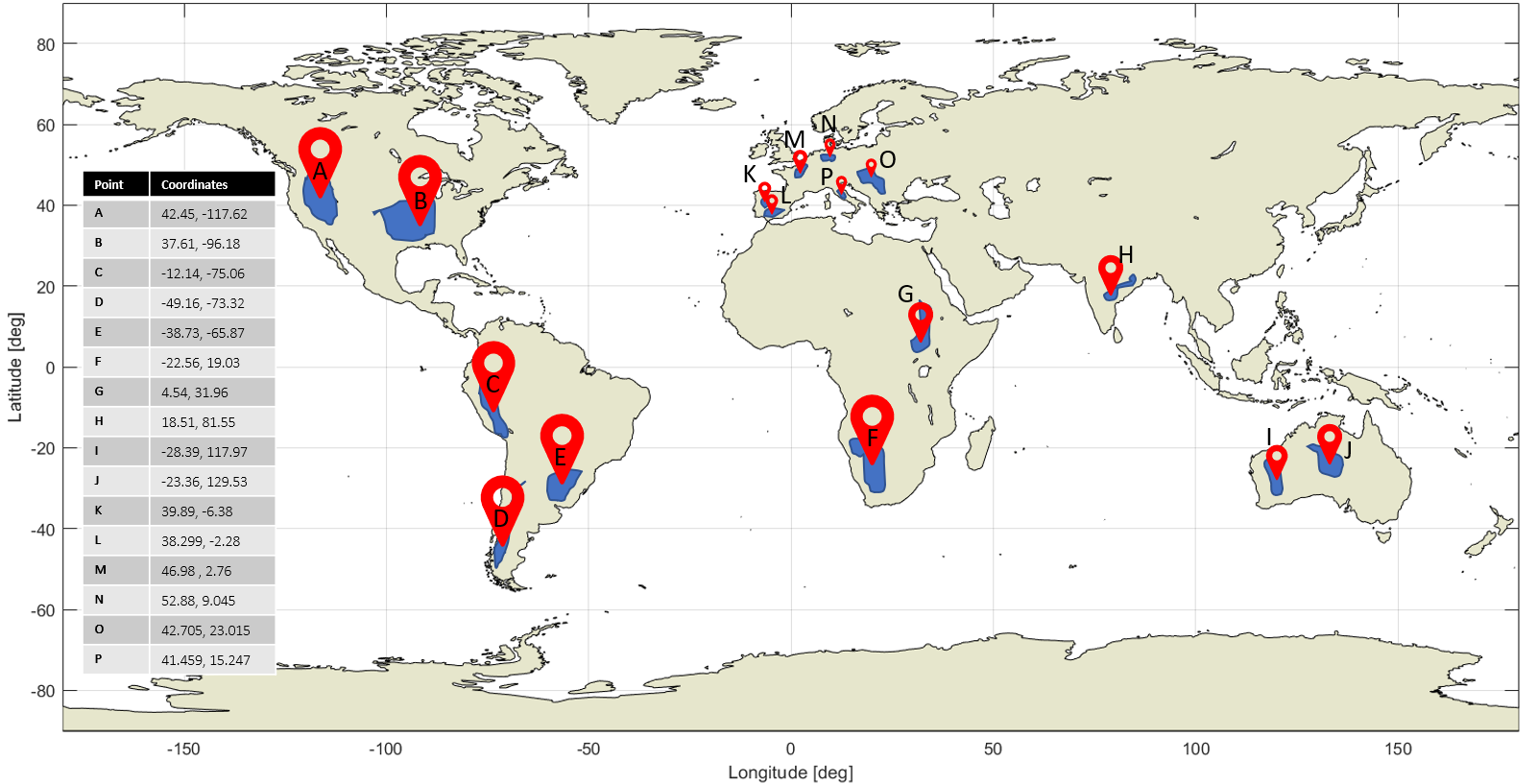} 
   \caption{World GW Positioning for NGSO system at 800 Km.}
   \label{positioning}
\end{figure*}

We have calculated how many positions with respect to the total pairs of coordinates that make up the grid exceed the thresholds established for each criterion.  The percentage of positions selected for each criterion is represented in \reff{results}.  We follow two cases to select the criteria: on the one hand, we are going to carry out a comparison using two by two grids; that is, we compare by pairs, where only two criteria are analyzed simultaneously. On the other hand, we analyze the general case in which all the selected criteria are considered simultaneously. We can note some criteria have a greater influence on others. The main influence is due to the chosen threshold, which is not optimized. Due to the length of the step between coordinates, we do not obtain exact and isolated positions in each region, but an area with multiple adjacent positions has similar characteristics, and therefore multiple GW in the same region are possible. Among all the GW of a region, we select the one that is in the position that coincides with the geographic mean of that area. Finally, in \reff{positioning}, the areas resulting from applying the proposed approach and the position for each GW selected in each of them are represented.

\begin{figure}[]%
    \centering
    \includegraphics[width=3.4in]{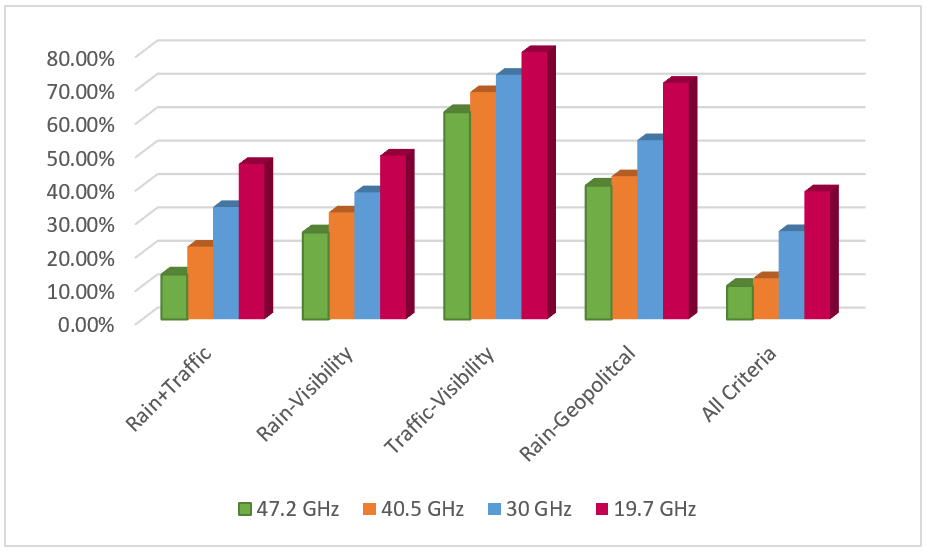} 
    \caption{Percentage of candidate positions to place GW.}
    \label{results}
\end{figure} 

\section{Discussion and Future Trends}
\label{research}

\subsection{Additional criteria} 

As mentioned in Section \ref{system}, additional criteria can be added as new grids to the proposed model, making it a very flexible tool. Below we provide a list of potential criteria that could be considered to enhance the current approach:

\subsubsection{\textbf{Elevation Angle (EA)}} EA is an important variable to consider when placing GW, as it can significantly affect signal propagation and link budget\footnote{Procedure for determining the received power, which ensures that the information is received intelligibly with an adequate signal-to-noise ratio.}. This parameter refers to the angle at which GW points to the satellite, which in turn is related to the height of the GW location above sea level. In this work, we have considered a fixed elevation for all coordinates.
The higher the EA of the GW, the higher the line-of-sight and link distance available for the signal, which can improve constellation performance. However, EA can also increase signal attenuation due to rain and other weather phenomena, which can negatively affect the quality of service. Therefore, choosing the appropriate GW elevation is a trade-off between performance, service quality, and attenuation due to rain. The methodology for placement should carefully consider the elevation and use advanced modeling and simulation techniques to determine the optimal height at each location. In addition, the methodology should be flexible enough to adapt to different meteorological and geographical conditions in different regions of the world.

\subsubsection{\textbf{Spectrum constraints}} Countries have different spectrum allocations for SatCom, which can limit the number of satellites that can be deployed and the frequencies they can use. Environmental and safety regulations must be considered, as well as restrictions on the location of the ground stations used to control and communicate with the satellites.

\subsubsection{\textbf{Regulatory constraints}} Deploying NGSO constellations requires a thorough understanding of the policies and regulations in each country and region. It must be done in consultation with the relevant regulators and authorities to ensure compliance with applicable regulations and policies. In addition, the geo-political policy may also impact the economics and cost of constellation deployment, as there may be taxes or tariffs associated with using space and frequencies in certain countries or regions. 

\subsubsection{\textbf{GW-Core distance}} Refers to the location of the GW in relation to the core of the terrestrial network. The network core is the central part of the communications network that processes and routes data through the network. The distance between the GW and the network core can affect system performance, especially end-to-end latency and network throughput.

To ensure optimal system performance, the methodology must consider the location of the GW relative to the network core. If the ground station or GW is too far from the network core, there may be excessive latency in data transmission, affecting service quality. On the other hand, if the GW is too close to the network core, depending on the network hierarchy\footnote{The network architecture has several layers organized hierarchically. This applies in space to multi-orbital constellations of different types.}, there may be congestion in data traffic that negatively affects performance.


\subsubsection{\textbf{Existing infrastructures integration}} A need exists to ensure that the mega-constellation integrates effectively and efficiently with existing infrastructure, such as other ground stations (teleports, hubs, VSAT terminals) and backhaul connections, which have to be prioritized if some GW is re-used in the planning. Proper integration is crucial to ensure continuity of service and signal quality. Interference between mega-constellation and terrestrial infrastructure, which could lead to the degradation of service quality and negatively affect the user experience, must be avoided. In addition, the availability of infrastructure resources for the mega-constellation must be ensured, which may require optimizing the use of resources and implementing new connectivity solutions.


\subsection{GW equipment}
The GW antenna is the main hardware element that can influence the number of GW needed and their positioning. Different types of antennas are used for gateways: symmetric or prime focus antennas, offset antennas, array antennas, and lens antennas. Among the symmetric and offset antennas, there are some variations using sub-reflectors to obtain antennas with our blockage of the feeder, for instance, Cassegrain or Gregorian, and even multi-frequency operation without defocusing the beams using dichroic sub-reflectors. Suppose the requirements of the antenna involve linking with multiple satellites at the same time with the same antenna infrastructure. In that case, multiple feeders aligned according to the constellation orbit can be cost-efficient.
On the other hand, antenna arrays used by GW for communication with the satellite display some drawbacks. For instance, grating lobes in an array of horn antennas can be a problem, with high production costs and high losses in the case of planar antennas. In addition, the authors in \cite{antenas} discuss antenna array integration concepts for combined receiver and transmitter terminals that can be used for ground-based Ka-band SATCOM. For Q/V/W bands and optical GW have to be extended. Lens antennas offer high bandwidth and, depending on the size, can generate highly focused beams. However, the biggest drawback of these antennas is the losses caused by the dielectric at high frequencies and its size. 
The design of the gateway antenna depends on the gain needed and the available transmission power, which can be calculated in the link budget calculation based on the satellite altitude, component losses, miss-match losses, antenna pointing loss, and propagation attenuation, just to mention the most relevant factors. For example, the reflector-based antenna can be used.

\subsection{Accurate weather modeling}

At the Q/V- bands, atmospheric attenuation can cause decades of dB magnitude losses, which can be even higher depending on the EA. Regarding the rain attenuation model, \cite{arquitectura} investigated the cumulative statistics of total attenuation induced in LEO mega-constellations operating at Q/V bands. The work in \cite{cornejo2022method}, as many others in this area, considers the ITU recommendation P.618-13 for the calculation of the exceeding probability of total attenuation for a given EA. These models have to be extended and improved to include the new spectral bands, such as Q/V and W.

\subsection{Multi-criteria optimization}

A multi-criteria optimization approach is required to optimize the placement of GW stations for NGSO satellites. This involves considering various technical and operational factors and constraints, such as traffic demand, rain attenuation, visibility time, geographic location, regulatory requirements, transmission power requirements, gateway processing and storage capacity, and integration with existing infrastructure. The weighting of all factors leads to a complex multi-objective optimization. Advanced mathematical models and optimization techniques, such as mathematical programming, genetic algorithms, and artificial neural networks, are the key to addressing these factors' complex and often conflicting requirements.  

Candidate methods for solving these optimization problems are classified into analytical optimization, metaheuristic optimization, and machine learning. Analytical optimization provides an optimal or near-optimal solution but may be time-consuming due to the large number of optimization parameters, making it unsuitable for real-time systems. Metaheuristic optimization, on the other hand, may not guarantee optimality but is well-suited for nonlinear, multi-objective, and hard problems. Machine learning algorithms interact with the environment or data to predict or decide on possible solutions, requiring less computational time than metaheuristic methods. However, they may not guarantee optimal solutions.

A trade-off often exists between performance and computational complexity. An example is when formulating the optimization results in a non-convex problem depending on the parameters chosen (capacity, traffic demand, latency, among others). This is often the case due to binary assignment variables, nonlinear expressions, and conflicting optimization variables. Therefore, optimizing the placement of GW stations for NGSO satellites is challenging and requires advanced modeling techniques, interdisciplinary expertise, and optimization methods that balance performance and computational complexity. In addition, the thresholds for each criteria can be optimized to consider different weightings depending on the design needs per operator.

\subsection{Inter-Satellite Links and space-routing}
The joint optimization or definition of the GS planning together with the Inter-Satellite Links (ISL) network can further reduce the number of elements needed both on the ground (number of GWs) and the number of hops between satellites in an ISL network. Such a design can potentially reduce deployment costs (estimated 70\% \cite{Hayder_Survey}), improve the security of services, reduce latency in communications, and facilitate the integration of future 6G. In addition, low-cost GS is a big opportunity for small operators to enter the new space age with NGSO constellations.

Associated with the joint optimization GS-ISL, it is necessary to develop routing\footnote{Paths of information between the different spatial elements that make up the constellation.} algorithms for multi-layer NGSO constellations: to address routing strategies, including multi-layer and multi-orbit NGSO/GEO architectures for the proliferation of large NGSO constellations equivalent to large graphs. 

From the perspective of the regulators, they demand to know the physical path that the information will follow between two GWs when it passes through ISL to ensure control, security, and management of the information. Extending the GW positioning network for worldwide coverage using ISL requires that a GW not necessarily have to be in the owner's country. Therefore, the regulation of this conjunction is a major concern. 

\begin{figure}[]%
    \centering
    \includegraphics[width=3.3in]{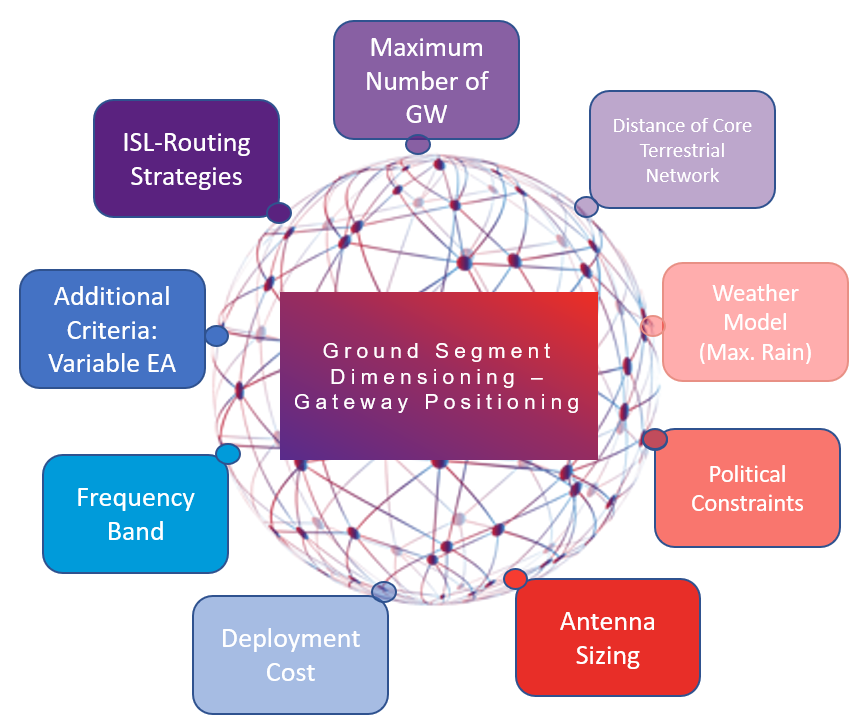} 
    \caption{Trade-off for GW positioning.}
    \label{tradeoffFig}
\end{figure} 

\subsection{Trade-off of different criteria}

The trade-off among different criteria to select the number of GW is crucial and evident to obtain an optimal ground segment for NGSO. Add to this the antenna size, the ISL network, the spectral restrictions, or new additional criteria, all interconnected as represented in \reff{tradeoffFig} to offer the best positions for the GWs. As an example, there is a clear trade-off between the number of GW (and total feeder link capacity) and the sizing antenna (which determines the single feeder link capacity). In general, assuming a full coverage of the constellation, smaller (antenna) GWs can be replaced by bigger (antenna) GW. A huge dish antenna GW can multiply the capacity of a small dish antenna GW. Therefore, multiple co-located GW can be replaced by a single ‘big’ one. 
Feeder links of future systems will most probably operate in Ka and Q/V. Therefore, those are the bands we propose as a priority. Within these frequency bands, Ka-band is at the moment more mature in terms of hardware (HW) components and requires less need for GW diversity. In addition, the deployment cost can limit the maximum number of GW. The EA is associated with the need to include or not diversity or overcome the attenuation imposed by temporal models, rain, or even cloud if we have optical GW.

On the other hand, assuming a high capacity and fully-connected ISL network may allow us to reduce in a significant manner the number of GWs and help in establishing them in secure locations.
Existing works mainly formulate the GW positioning problem as an integer linear programming, with the objectives typically being the number of GW minimization or maximization of network capacity. Adding the implementation costs to the problem formulation as well as the specific hardware equipment, may render too many degrees of freedom into the problem, which are most of the times intertwined.

\section{Conclusions}
\label{sec:conclusions}
In this work, we presented an overview of the key factors to consider for NGSO gateway station positioning, proposed a ground segment dimensioning approach that combines several criteria, and discussed a case study demonstrating the performance of the proposed methodology for one sample constellation. The approach is presented from an operational perspective. The paper concludes with a discussion of relevant open research challenges and potential research directions.

\section*{Acknowledgements}
\label{sec:ack}


This work has been supported by the project TRANTOR, which has received funding from the European Union’s Horizon Europe research and innovation program under grant agreement No. 101081983.

\balance
\bibliographystyle{IEEEtran}
\bibliography{references}

\vfill 

VICTOR MONZON BAEZA received the M.Sc. and Ph.D. degrees in Electrical Engineering from the University Carlos III of Madrid, Spain, 2013 and 2019, respectively. Currently, he is a research associate at the University of Luxembourg, Luxembourg. His main research interests include signal-processing and optimizing non-coherent massive-MIMO and satellite communications.\\  

FLOR ORTIZ 
received the M.Sc. and Ph.D. degrees in Electrical Engineering from Universidad Nacional Autónoma de México, Mexico, in 2016 and Universidad Politécnica de Madrid, Spain, in 2021, respectively. Currently, she is a research associate at the University of Luxembourg, Luxembourg. Her research interests include machine learning in satellite communication.\\ 

EVA LAGUNAS
received the M.Sc. and Ph.D. degrees in Telecommunications Engineering from the Polytechnic University of Catalonia (UPC), Barcelona, Spain, in 2010 and 2014, respectively. Currently, she is a research scientist at the University of Luxembourg, Luxembourg. Her research interests include radio resource management and general wireless network optimization.
\\ 

TEDROS SALIH ABDU
received the Ph.D. degree in Computer Science from the University of Luxembourg, Luxembourg, in 2022. Currently, he is a research associate at the University of Luxembourg, Luxembourg. His main research interests include radio resource management in satellite communications.  \\ 

SYMEON CHATZINOTAS  
received the Ph.D. degree in electronic engineering from the University of Surrey, U.K., in 2009. He is currently a Full Professor/Chief Scientist I and Head of the SIGCOM Research Group with SnT, University of Luxembourg.  He is the main representative for 3GPP, ETSI, DVB.

\end{document}